\documentstyle[amssymb,11pt]{article}

\newlength{\minitwocolumn}
\makeatletter
\@addtoreset{equation}{section}
\makeatother

\newtheorem{thm}{Theorem}[section]
\newtheorem{prop}[thm]{Proposition}

\newtheorem{df}{Definition}[section]
\newtheorem{dfn}[thm]{Definition}
\title{\bf
\huge{\bf
Feigin-Odesskii algebra and its realization
in ellptic deformation of field theory}}
\begin{document}
~\\
\begin{Large}
{\bf
Free field realization of
commutative family of \\
elliptic Feigin-Odesskii algebra
}
\end{Large}
\\
~\\
~~~~~~~~~~~~~~~~~~~~~~~~~~~~~~~~~~~~~~~~~
%日本大学・理工学部・数学科　　小島　武夫
~~~~~~~~~~~~~~~~~~~~~~~~~~~~~~~~~~~~~Takeo Kojima\\
~~~~~~~~~~~~~~~~~~~~~~~~~~~~~~~~~~~~~~~~~~~~
~~~~~~~~~~~~~~~~~~~~~~~~~~~~~~~~~~Department of Mathematics,\\
~~~~~~~~~~~~~~~~~~~~~~~~~~~~~~~~~
~~~~~~~~~~~~~~~~~~~~~~~~~~~~~~~~~~~~~~~~~~~~~College of Science and Technology,\\
~~~~~~~
~~~~~~~~~~~~~~~~~~~~~~~~~~~~~~~~~~~
~~~~~~~~~~~~~~~~~~~~~~~~~~~~~~~~~~~Nihon University\\

\begin{center}
{\bf Abstract}
\end{center}
In this review, we study
free field realizations of
the Feigin-Odesskii algebra.
We construct free field realizations
of a pair of infinitely many commutative operators,
associated with the elliptic algebra $U_{q,p}(\widehat{sl_N})$.

\section{Introduction}

In this review, we study
free field realization of
elliptic version of the Feigin-Odesskii algebra
\cite{FO}.
For this purpose we introduce one parameter "$s$"
deformation of the Feigin-Odesskii algebra \cite{FO}.
This review is based on the paper
\cite{FKSW1,KS1,KS2,K1,K2}.
Let the function $f_l(z_1\cdots z_l|w_1\cdots w_l)$
be meromorphic and symmetric in each of varibles
$(z_1,\cdots,z_l)$ and $(w_1,\cdots,w_l)$.
Let us set the symmetric function
$(f_m \circ f_n)(z_1,\cdots,z_{m+n}|w_1,\cdots,w_{m+n})$,
depending on three
continuous parameters $0<x<1,0<r$ and $0<s<2$, by
\begin{eqnarray}
&&(f_m \circ f_n)(z_1,\cdots,z_{m+n}|
w_1,\cdots,w_{m+n})\nonumber\\
&=&
\frac{1}{((m+n)!)^2}
\sum_{\sigma \in S_{m+n}}
\sum_{\tau \in S_{m+n}}
f_m(z_{\sigma(1)},\cdots,z_{\sigma(m)}|
w_{\tau(1)},\cdots,w_{\tau(m)})\nonumber\\
&&\times
f_n(z_{\sigma(m+1)},\cdots,z_{\sigma(m+n)}|
w_{\tau(m+1)},\cdots,w_{\tau(m+n)})
\nonumber
%\label{def:FO1}
\\
&&\times
\prod_{i=1}^m \prod_{j=m+1}^{m+n}
\frac
{\displaystyle
\left[v_{\tau(i)}-u_{\sigma(j)}+\frac{s}{2}\right]_r
\left[u_{\sigma(i)}-v_{\tau(j)}+\frac{s}{2}\right]_r
}
{
\displaystyle
[u_{\sigma(i)}-u_{\sigma(j)}]_{r}
[u_{\sigma(j)}-u_{\sigma(i)}-1]_r
}.\nonumber
\\
&&\times
\prod_{i=1}^m \prod_{j=m+1}^{m+n}
\frac
{\displaystyle
\left[u_{\sigma(j)}-v_{\tau(i)}+\frac{s}{2}-1\right]_r
\left[v_{\tau(j)}-u_{\sigma(i)}+\frac{s}{2}-1\right]_r}
{
\displaystyle
[v_{\tau(j)}-v_{\tau(i)}-1]_r
[v_{\tau(j)}-v_{\tau(i)}-1]_r},\nonumber
\end{eqnarray}
where the symbol $[u]_r$ represents
the elliptic theta function defined in (\ref{def:theta}).
Here we set $z_j=x^{2u_j}, w_j=x^{2v_j}$.
This product "$\circ$" on symmetric function gives
the structure of the associative algebra.
We call this associative algebra 
"elliptic Feigin-Odesskii algebra".
Let us set the functional ${\cal G}$ by using
currents $F_1(z),F_2(z)$,
which is one parameter "$s$" deformation of
the elliptic algebra $U_{q,p}(\widehat{sl_2})$.
They satisfy the following commutation relations.
\begin{eqnarray}
\left[u_1-u_2-\frac{s}{2}\right]_r
\left[u_1-u_2+\frac{s}{2}-1\right]_r
F_1(z_1)F_2(z_2)
&=&
\left[u_2-u_1-\frac{s}{2}\right]_r
\left[u_2-u_1+\frac{s}{2}-1\right]_r
F_1(z_2)F_2(z_1),\nonumber\\
~[u_1-u_2]_r[u_1-u_2+1]_r
F_j(z_1)F_j(z_2)
&=&
[u_2-u_1]_r[u_2-u_1+1]_r
F_j(z_2)F_j(z_1).\nonumber
\end{eqnarray}
Upon the specialization $s \to 2$
the current $F_1(z)$ degenerates to
the cuurent of the elliptic algebra $U_{q,p}(\widehat{sl_2})$,
and the current $F_2(z)$ lookes like $F_1(z)^{-1}$.
Let us set the functional ${\cal G}$ by
\begin{eqnarray}
{\cal G}(f_m)
&=&
\oint \prod_{j=1}^m\frac{dz_j}{2\pi i z_j}
\oint \prod_{j=1}^m\frac{dw_j}{2\pi i w_j}
F_1(z_1)\cdots F_1(z_m)F_2(w_1)\cdots F_2(w_m)
\nonumber
%\label{def:functional1}
\\
&\times&
\frac{\displaystyle
\prod_{1\leqq j<k \leqq m}
[u_i-u_j]_r[u_j-u_i-1]_r
[v_i-v_j]_r[v_j-v_i-1]_r}{
\displaystyle
\prod_{i=1}^m
\prod_{j=1}^m
\left[u_i-v_j+\frac{s}{2}\right]_r
\left[v_j-u_i+\frac{s}{2}-1\right]_r
}f_m(z_1,\cdots,z_m|w_1,\cdots,w_m).\nonumber
\end{eqnarray}
Roghly speaking, this functional satisfies homomorphism,
\begin{eqnarray}
{\cal G}(f_m){\cal G}(f_n)=
{\cal G}(f_m \circ f_n),
\nonumber%\label{eqn:functional1}
\end{eqnarray}
which is a consequence of
symmetrizing procedure of variables $(z_1,\cdots,z_{m+n})$
and $(w_1,\cdots,w_{m+n})$.
We call ${\cal G}(f_m)$ 
"free field realization of Feigin-Odesskii algebra".
When we have commutative family
$\vartheta_m$ of elliptic Feigin-Odesskii algebra,
\begin{eqnarray}
\vartheta_m \circ \vartheta_n=
\vartheta_n \circ \vartheta_m,\nonumber
\end{eqnarray}
we can construct
commutative family of the operators ${\cal G}(\vartheta_m)$,
\begin{eqnarray}
{\cal G}(\vartheta_m)\cdot {\cal G}(\vartheta_n)=
{\cal G}(\vartheta_n)\cdot {\cal G}(\vartheta_m).\nonumber
\end{eqnarray}
This is rough story of this paper.
Precisely this homomorphism ${\cal G}(f_m){\cal G}(f_n)=
{\cal G}(f_m \circ f_n)$
does not hold for every time.
For example, upon the specialization
$s\to 2$,
the homomorphism does not hold, 
because singularity which comes from
the product of the current $E_1(z), E_2(w)$,
destroy the structure.
In order to construct the free field realization of
Feigin-Odesskii algebra, we have to construct the currents
which 
(1) satisfy the commutation relation, and
(2) does not have surplus singularity.
In this survey we construct free field realization
of commuttive family of
elliptic Feigin-Odesskii algebra.

The organization of this paper is as follows.
In section 2 
we introduce a pair of Feigin-Odesskii algebra,
and give infinitely many commutative solutions of
the Feigin-Odesskii algebra.
We construct free field realization of
the Feigin-Odesskii algebra, by using
one parameter deformation of
the current of 
the elliptic algebra $U_{q,p}(\widehat{sl_2})$.
In terms of this free field realization,
we construct
a pair of
infinitely many commutative 
operators acting on the Fock space.
In section 3 
we consider the higher-rank generalization of section 2.
We construct a pair infinitely many commutative 
operators by using one parameter deformation of
the elliptic algebra 
$U_{q,p}(\widehat{sl_N})$.
In section 4 
we consider higher-level $k$ generalization of
section 2.
We construct free field realization of
one parameter deformation of level $k$ elliptic algebra
$U_{q,p}(\widehat{sl_2})$.
The author would like to
emphasize that
the free field realization of Level $k$ is completely 
different from those of Level $1$.
We construct
a pair of 
infinitely many commutative operators associated with 
one parameter $s$ deformation of
the elliptic algebra 
$U_{q,p}(\widehat{sl_2})$ for level $k$.
In section 5 we
give a free field realization of the elliptic algebra
$U_{q,p}(\widehat{sl_N})$ for level $k$, and
explain an open problem.
In section 2 we summarize some of results in \cite{FKSW1}.
In section 3 we summarize some of results in
\cite{KS1}.
In section 4 we summarize the results
in \cite{KS2}.
In section 5 we summarize the results in
\cite{K1,K2}.

\section{Elliptic algebra $U_{q,p}(\widehat{sl_2})$}

Let us fix parameters $0<x<1$, $r>0$.
Let us set $z=x^{2u}$.
The symbol $[u]_r$ stands for the Jacobi theta function,
\begin{eqnarray}
[u]_r=x^{\frac{u^2}{r}-u}\frac{\Theta_{x^{2r}}(z)}{
(x^{2r};x^{2r})_\infty^3},
~~\Theta_q(z)=
(q;q)_\infty
(z;q)_\infty
(q/z;q)_\infty,
\label{def:theta}
\end{eqnarray}
where we have used standard notation $(z;q)_\infty
=\prod_{j=0}^\infty (1-q^jz)$.
The symbol $[a]$ stands for $q$-integer,
\begin{eqnarray}
[a]=\frac{x^a-x^{-a}}{x-x^{-1}}.
\end{eqnarray}

\subsection{Feigin-Odesskii algebra}

Let us set parameters $0<s<2$ and $r>1$.
We introduce a pair of Feigin-Odesskii algebra:
$f \circ g$ and $f * g$.

\begin{dfn}~~
Let us set the symmetric function 
$(f_m \circ f_n)(z_1,\cdots,z_{m+n}|w_1,\cdots,w_{m+n})$ 
by
\begin{eqnarray}
&&(f_m \circ f_n)(z_1,\cdots,z_{m+n}|
w_1,\cdots,w_{m+n})\nonumber\\
&=&
\frac{1}{((m+n)!)^2}
\sum_{\sigma \in S_{m+n}}
\sum_{\tau \in S_{m+n}}
f_m(z_{\sigma(1)},\cdots,z_{\sigma(m)}|
w_{\tau(1)},\cdots,w_{\tau(m)})\nonumber\\
&&\times
f_n(z_{\sigma(m+1)},\cdots,z_{\sigma(m+n)}|
w_{\tau(m+1)},\cdots,w_{\tau(m+n)})
\nonumber
\\
&&\times
\prod_{i=1}^m \prod_{j=m+1}^{m+n}
\frac
{\displaystyle
\left[v_{\tau(i)}-u_{\sigma(j)}+\frac{s}{2}\right]_r
\left[u_{\sigma(i)}-v_{\tau(j)}+\frac{s}{2}\right]_r
}
{
\displaystyle
[u_{\sigma(i)}-u_{\sigma(j)}]_{r}
[u_{\sigma(j)}-u_{\sigma(i)}-1]_r
}\nonumber\\
&\times&
\prod_{i=1}^m \prod_{j=m+1}^{m+n}
\frac
{\displaystyle
\left[u_{\sigma(j)}-v_{\tau(i)}+\frac{s}{2}-1\right]_r
\left[v_{\tau(j)}-u_{\sigma(i)}+\frac{s}{2}-1\right]_r}
{
\displaystyle
[v_{\tau(j)}-v_{\tau(i)}-1]_r
[v_{\tau(j)}-v_{\tau(i)}-1]_r}.\label{def:FO1}
\end{eqnarray}
Let us give the symmetric function
$(f_m * f_n)(z_1 \cdots z_{m+n}|w_1 \cdots w_{m+n})$ by 
\begin{eqnarray}
&&(f_m * f_n)(z_1,\cdots,z_{m+n}|
w_1,\cdots,w_{m+n})\nonumber\\
&=&
\frac{1}{((m+n)!)^2}
\sum_{\sigma \in S_{m+n}}
\sum_{\tau \in S_{m+n}}
f_m(z_{\sigma(1)},\cdots,z_{\sigma(m)}|
w_{\tau(1)},\cdots,w_{\tau(m)})\nonumber\\
&&\times
f_n(z_{\sigma(m+1)},\cdots,z_{\sigma(m+n)}|
w_{\tau(m+1)},\cdots,w_{\tau(m+n)})\nonumber
\\
&&\times
\prod_{i=1}^m \prod_{j=m+1}^{m+n}
\frac
{\displaystyle
\left[v_{\tau(i)}-u_{\sigma(j)}-\frac{s}{2}\right]_{r-1}
\left[u_{\sigma(i)}-v_{\tau(j)}-\frac{s}{2}\right]_{r-1}
}
{
\displaystyle
[u_{\sigma(i)}-u_{\sigma(j)}]_{r-1}
[u_{\sigma(j)}-u_{\sigma(i)}+1]_{r-1}
}\nonumber\\
&\times&
\prod_{i=1}^m \prod_{j=m+1}^{m+n}
\frac
{\displaystyle
\left[u_{\sigma(j)}-v_{\tau(i)}-\frac{s}{2}+1\right]_{r-1}
\left[v_{\tau(j)}-u_{\sigma(i)}-\frac{s}{2}+1\right]_{r-1}}
{
\displaystyle
[v_{\tau(j)}-v_{\tau(i)}+1]_{r-1}
[v_{\tau(j)}-v_{\tau(i)}+1]_{r-1}}.\label{def:FO2}
\end{eqnarray}
Here $f_l(z_1,\cdots,z_l|w_1,\cdots,w_l)$
are meromorphic function symmetric in each of varibles
$(z_1,\cdots,z_l)$ and $(w_1,\cdots,w_l)$.
\end{dfn}

We have infinitely mny commutative 
family of Feigin-Odesskii algebra.
Let us set theta functions for 
three parameters $\alpha, \nu$,
\begin{eqnarray}
&&\vartheta_{m,\alpha}
(z_1,\cdots,z_m|w_1,\cdots,w_m)=
\left[\sum_{j=1}^m(u_j-v_j)-\nu+\alpha \right]_r
\left[\sum_{j=1}^m(v_j-u_j)-\alpha \right]_r.
\label{def:theta1}
\end{eqnarray}

\begin{prop}~~$\vartheta_{m,\alpha}$ 
and $\vartheta_{n,\beta}$ commute
with respect to the product (\ref{def:FO1}).
\begin{eqnarray}
\vartheta_{m,\alpha} \circ \vartheta_{n,\beta}
=\vartheta_{n,\beta} \circ \vartheta_{m,\alpha}.
\end{eqnarray}
\end{prop}

Let us set theta functions for parmeters $\alpha, \nu$.
\begin{eqnarray}
&&\vartheta^*_{m,\alpha}
(z_1,\cdots,z_m|w_1,\cdots,w_m)=
\left[\sum_{j=1}^m(v_j-u_j)-\nu+\alpha \right]_{r-1}
\left[\sum_{j=1}^m(u_j-v_j)-\alpha \right]_{r-1}.
\label{def:theta2}
\end{eqnarray}

\begin{prop}~~$\vartheta_{m,\alpha}^*$ and 
$\vartheta_{n,\beta}^*$ commute
with respect to the product (\ref{def:FO2}).
\begin{eqnarray}
\vartheta_{m,\alpha}^* * \vartheta_{n,\beta}^*
=\vartheta_{n,\beta}^* * \vartheta_{m,\alpha}^*.
\end{eqnarray}
\end{prop}
Proof of propositions are summarized in \cite{FKSW1}.

\subsection{Free field realization}

Let us set a parameter $0<s<2$.
Let us introduce bosons 
$\beta_m^1,\beta_m^2$, $(m\neq 0)$
by
\begin{eqnarray}
[\beta_m^i,\beta_n^j]=\left\{
\begin{array}{cc}
m\frac{[(r-1)m]}{[rm]}\frac{[(s-1)m]}{[sm]}\delta_{m+n,0},&
(i=j)\\
-m\frac{[(r-1)m]}{[rm]}\frac{[m]}{[sm]}x^{sm~{\rm sgn}(i-j)}
\delta_{m,n},& (i\neq j)
\end{array}
\right.
\end{eqnarray}
Let us set $P,Q$ by
\begin{eqnarray}
&&~[P,iQ]=1.
\end{eqnarray}
We deal with
the bosonic Fock space
${\cal F}_{l,k}$,$(l,k \in {\mathbb Z})$
generated by
$\beta_{-m}^i$,$(m>0, i=1,2)$
over the vacuum vector $|l,k\rangle$.
\begin{eqnarray}
&&~\beta_m^i |l,k\rangle=0~(m>0,i=1,2),\\
&&~P|l,k\rangle=
\left(
\sqrt{\frac{r}{2(r-1)}}l-\sqrt{\frac{r-1}{2r}}k
\right)|l,k\rangle,\\
&&|l,k\rangle=
e^{\left(
\sqrt{\frac{r}{2(r-1)}}l-\sqrt{\frac{r-1}{2r}}k
\right)iQ}|0,0\rangle.
\end{eqnarray}

\begin{dfn}~~~
Let us set the currents $F_j(z),E_j(z)$,$(j=1,2)$ by
\begin{eqnarray}
F_1(z)&=&z^{\frac{r-1}{r}}
e^{i\sqrt{\frac{2(r-1)}{r}}Q}
z^{\sqrt{\frac{2(r-1)}{r}}P}
:\exp\left(\sum_{m\neq 0}\frac{1}{m}
(\beta_m^1-\beta_m^2)z^{-m}
\right):,\\
F_2(z)&=&
z^{\frac{r-1}{r}}
e^{-i\sqrt{\frac{2(r-1)}{r}}Q}
z^{-\sqrt{\frac{2(r-1)}{r}}P}
:\exp\left(\sum_{m\neq 0}\frac{1}{m}
(-x^{sm}\beta_m^1+x^{-sm}
\beta_m^2)z^{-m}
\right):,\\
E_1(z)&=&z^{\frac{r}{r-1}}
e^{-i\sqrt{\frac{2r}{r-1}}Q}
z^{-\sqrt{\frac{2r}{r-1}}P}
:\exp\left(-\sum_{m\neq 0}\frac{1}{m}
\frac{[rm]}{[(r-1)m]}(\beta_m^1-\beta_m^2)z^{-m}
\right):,\\
E_2(z)&=&
z^{\frac{r}{r-1}}
e^{i\sqrt{\frac{2r}{r-1}}Q}
z^{\sqrt{\frac{2r}{r-1}}P}
:\exp\left(-\sum_{m\neq 0}\frac{1}{m}
\frac{[rm]_x}{[(r-1)m]_x}(-x^{sm}\beta_m^1+x^{-sm}
\beta_m^2)z^{-m}
\right):.
\end{eqnarray}
\end{dfn}
They satisfy the following commutation relations.

\begin{prop}~~
\begin{eqnarray}
\frac{[u_1-u_2]_r}{[u_1-u_2-1]_r}
F_j(z_1)F_j(z_2)&=&
\frac{[u_2-u_1]_r}{[u_2-u_1-1]_r}
F_j(z_2)F_j(z_1),~~(j=1,2)\\
\frac{[u_1-u_2+\frac{s}{2}-1]_r}
{[u_1-u_2+\frac{s}{2}]_r}
F_1(z_1)F_2(z_2)&=&
\frac{[u_2-u_1+\frac{s}{2}-1]_r}{[u_2-u_1+\frac{s}{2}]_r}
F_2(z_2)F_1(z_1),\\
%%%%%%%%%%%%%%%%%%%%%%%%%%%%%%%%%
\frac{[u_1-u_2]_{r-1}}{[u_1-u_2+1]_{r-1}}
E_j(z_1)E_j(z_2)&=&
\frac{[u_2-u_1]_{r-1}}{[u_2-u_1+1]_{r-1}}
E_j(z_2)E_j(z_1),~~(j=1,2)\\
\frac{[u_1-u_2-\frac{s}{2}+1]_{r-1}}
{[u_1-u_2-\frac{s}{2}]_{r-1}}
E_1(z_1)E_2(z_2)&=&
\frac{[u_2-u_1-\frac{s}{2}+1]_{r-1}}
{[u_2-u_1-\frac{s}{2}]_{r-1}}
E_2(z_2)E_1(z_1).
\end{eqnarray}
\begin{eqnarray}
&&[E_i(z_1),F_j(z_2)]\nonumber\\
&=&
\frac{\delta_{i,j}}{x-x^{-1}}
\left(\delta(xz_2/z_1)H_j(x^rz_2)-
\delta(xz_1/z_2)H_j(x^{-r}z_2)\right),~~(i,j=1,2).
\end{eqnarray}
Here we have set
\begin{eqnarray}
H_1(z)&=&e^{-\frac{1}{\sqrt{r (r-1)}}iQ}
z^{-\frac{1}{\sqrt{r (r-1)}}P+\frac{1}{r (r-1)}}
:\exp\left(-\sum_{m \neq 0}
\frac{1}{m}\frac{[m]}{[(r-1)m]}(\beta_m^1-\beta_m^2)
z^{-m}\right):,\\
H_2(z)&=&
e^{\frac{1}{\sqrt{r (r-1)}}iQ}
z^{\frac{1}{\sqrt{r (r-1)}}P+\frac{1}{r (r-1)}}
:\exp\left(\sum_{m \neq 0}
\frac{1}{m}\frac{[m]}{[(r-1)m]}(x^{sm}\beta_m^1-
x^{-sm}\beta_m^2)
z^{-m}\right):.
\end{eqnarray}
\end{prop}

\begin{dfn}~~
Let us set the functional ${\cal G}$ by
\begin{eqnarray}
{\cal G}(f_m)
&=&
\oint \prod_{j=1}^m\frac{dz_j}{2\pi i z_j}
\oint \prod_{j=1}^m\frac{dw_j}{2\pi i w_j}
F_1(z_1)\cdots F_1(z_m)F_2(w_1)\cdots F_2(w_m)
\label{def:functional1}\\
&\times&
\frac{\displaystyle
\prod_{1\leqq j<k \leqq m}
[u_i-u_j]_{r}[u_j-u_i-1]_{r}
[v_i-v_j]_{r}[v_j-v_i-1]_{r}}{
\displaystyle
\prod_{i=1}^m
\prod_{j=1}^m
\left[u_i-v_j+\frac{s}{2}\right]_{r}
\left[v_j-u_i-\frac{s}{2}+1\right]_{r^*}
}f_m(z_1,\cdots,z_m|w_1,\cdots,w_m).\nonumber
\end{eqnarray}
We take the integration contours to be simple closed curves
around the origin satisfying
\begin{eqnarray}
|x^sw_i|,|x^{2-s}w_i|<|z_j|<|x^{-s}w_i|,
|x^{s-2}w_i|,~~(i,j=1,2,\cdots,m).\nonumber
\end{eqnarray}
Let us set the functional
${\cal G}^*$ by followings.
\begin{eqnarray}
{\cal G}^*(f_m)
&=&
\oint \prod_{j=1}^m\frac{dz_j}{2\pi i z_j}
\oint \prod_{j=1}^m\frac{dw_j}{2\pi i w_j}
E_1(z_1)\cdots E_1(z_m)E_2(w_1)\cdots E_2(w_m)
\label{def:functional2}\\
&\times&
\frac{\displaystyle
\prod_{1\leqq j<k \leqq m}
[u_i-u_j]_{r-1}[u_j-u_i+1]_{r-1}
[v_i-v_j]_{r-1}[v_j-v_i+1]_{r-1}}{
\displaystyle
\prod_{i=1}^m
\prod_{j=1}^m
\left[u_i-v_j-\frac{s}{2}\right]_{r-1}
\left[v_j-u_i-\frac{s}{2}+1\right]_{r-1}
}f_m(z_1,\cdots,z_m|w_1,\cdots,w_m).\nonumber
\end{eqnarray}
We take the integration contours to be simple closed curves
around the origin satisfying
\begin{eqnarray}
|x^sw_i|,|x^{2-s}w_i|<|z_j|<|x^{-s}w_i|,
|x^{s-2}w_i|,~~(i,j=1,2,\cdots,m).\nonumber
\end{eqnarray}
\end{dfn}

\begin{prop}~~
When the function $f_l(z_1,\cdots,z_l|w_1,\cdots,w_l)$
are meromorphic function symmetric in each of varibles
$(z_1,\cdots,z_l)$, $(w_1,\cdots,w_l)$, and
don't have poles at the origin $z_j=0$, $w_j=0$,
the functionals ${\cal G}$, ${\cal G}^*$ satisfy
\begin{eqnarray}
&&{\cal G}(f_m){\cal G}(f_n)=
{\cal G}(f_m \circ f_n),\\
&&{\cal G}^*(f_m){\cal G^*}(f_n)=
{\cal G}^*(f_m*f_n).
\end{eqnarray}
\end{prop}
Symmetrizing with respect to the integration variables
$(z_1,\cdots,z_{m+n})$,
$(w_1,\cdots,w_{m+n})$ of 
product ${\cal G}(f_m){\cal G}(f_n)$, 
we have the above proposition.
We have to choose the integration contours symmetric
with respect with integration variables.
Hence, following normal orderings,
\begin{eqnarray}
F_1(z)F_2(w)=::x^{-\frac{2(r-1)}{r}}
\frac{
(x^{2r-2+s}w/z;x^{2r})_\infty 
(x^{2r-s}w/z)_\infty}{(
x^{s}w/z;x^{2r})_\infty
(x^{2-s}w/z;x^{2r})_\infty},\nonumber
\\
F_2(w)F_1(z)=::
x^{-\frac{2(r-1)}{r}}
\frac{
(x^{2r-2+s}z/w;x^{2r})_\infty 
(x^{2r-s}z/w)_\infty}{(
x^{s}z/w;x^{2r})_\infty
(x^{2-s}z/w;x^{2r})_\infty}.\nonumber
\end{eqnarray}
we have to choose the integration contours
to be simple closed curves
around the origin satisfying
\begin{eqnarray}
|x^sw_i|,|x^{2-s}w_i|<|z_j|<|x^{-s}w_i|,
|x^{s-2}w_i|,~~(i,j=1,2,\cdots,m).\nonumber
\end{eqnarray}
When we consider the case $s\to 2$ or $s\to 0$,
there does not exist such a contour.
Hence the above proposition does not hold.
We note
that one deformation parameter $0<s<2$ plays an essential role
in construction of commutative operators.
When we take the limit $s \to 2$, we get popular current
of the elliptic algebra $U_{q,p}(\widehat{sl_2})$,
and
the free field realization of
Feigin-Odesskii algebra is open problem.
In what follows, we set $\nu=\sqrt{r(r-1)}P$.

\begin{thm}~~For $r>1$ we have
\begin{eqnarray}
&&[{\cal G}(\vartheta_{m,\alpha}),
{\cal G}(\vartheta_{n,\beta})]=0,~~
(m,n \in {\mathbb N}),\\
&&[{\cal G}^*(\vartheta_{m,\alpha}^*),{\cal G}^*
(\vartheta^*_{n,\beta})]=0,~~
(m,n \in {\mathbb N}).
\end{eqnarray}
\end{thm}

\begin{thm}~~For $0<r<1$ we have
\begin{eqnarray}
[{\cal G}(\vartheta_{m,\alpha}),
{\cal G}^*(\vartheta^*_{n,\beta})]=0,~~
(m,n \in {\mathbb N}).
\end{eqnarray}
\end{thm} 
Definition of ${\cal G}^*(\vartheta^*_{m,\alpha})$
for $0<r<1$ is given as the same manner as 
(\ref{def:functional2}). See detailds in \cite{FKSW1}.
We have constructed infinitely many
commutative operators
${\cal G}(\vartheta_{m,\alpha})$, 
${\cal G}^*(\vartheta_{m,\alpha}^*)$,
$(m \in {\mathbb N})$ 
acting on the bosonic Fock space,
which is regarded as the free field realization
of commutative family of
Feigin-Odesskii algebra (\ref{def:FO1}) and (\ref{def:FO2}).

%%%%%%%%%%%%%%%%%%%%%%%%%%%%%%%%%%%%%%%%%%%%%%%%%%

\section{Elliptic algebra $U_{q,p}(\widehat{sl_N})$}

In this section we summarize
some of results in \cite{KS1}.
In this section 
we fix $N=3,4,\cdots$.
We set parameters $0<s<N$.

\subsection{Feigin-Odesskii algebra}

We introduce a pair of Feigin-Odesskii algebra.
We set $z_j^{(t)}=x^{2u_j^{(t)}}$ 
and understand $z_j^{(t+N)}=z_j^{(t)}$.

\begin{dfn}~~Let us set meromorphic function
$(f_m \circ f_n)(z_1^{(1)},\cdots, z_{m+n}^{(1)}|
\cdots|
z_1^{(N)},\cdots, z_{m+n}^{(N)})$
symmetric in each of variables
$(z_1^{(1)},\cdots, z_{m+n}^{(1)})$,$\cdots$
$(z_1^{(N)},\cdots, z_{m+n}^{(N)})$.
\begin{eqnarray}
&&(f_m \circ f_n)
(z_1^{(1)},\cdots,z_{m+n}^{(1)}|
\cdots|
z_1^{(N)},\cdots,z_{m+n}^{(N)})
\nonumber\\
&=&
\sum_{\sigma_1 \in S_{m+n}}
\sum_{\sigma_2 \in S_{m+n}}
\cdots
\sum_{\sigma_N \in S_{m+n}}\nonumber\\
&\times&f_m
(z_{\sigma_1(1)}^{(1)},\cdots,z_{\sigma_1(m)}^{(1)}|
\cdots|
z_{\sigma_N(1)}^{(N)},\cdots,
z_{\sigma_N(m)}^{(N)})
\nonumber\\
&\times&
f_n
(z_{\sigma_1(m+1)}^{(1)},\cdots,z_{\sigma_1(m+n)}^{(1)}|
\cdots|
z_{\sigma_N(m+1)}^{(N)},\cdots,
z_{\sigma_N(m+n)}^{(N)})
\nonumber\\
&\times&
\prod_{t=1}^N
\prod_{i=1}^m
\prod_{j=m+1}^{m+n}
\frac{\displaystyle
\left[
u_{\sigma_t(i)}^{(t)}-u_{\sigma_{t+1}(j)}^{(t+1)}-\frac{s}{N}
\right]_r
\left[
u_{\sigma_{t+1}(i)}^{(t+1)}-
u_{\sigma_{t}(j)}^{(t)}+1-\frac{s}{N}
\right]_r
}{
\displaystyle
\left[
u_{\sigma_t(i)}^{(t)}-u_{\sigma_{t}(j)}^{(t)}
\right]_r
\left[
u_{\sigma_{t}(j)}^{(t)}-
u_{\sigma_{t}(i)}^{(t)}-1\right]_r
}.\label{def:FO3}
\end{eqnarray}
Here meromorphic function
$f_l(z_1^{(1)},\cdots,z_l^{(1)}|
\cdots|
z_1^{(N)},\cdots,z_l^{(N)})$
is symmetric in each of variables
$(z_1^{(1)},\cdots,z_l^{(1)})$,
$\cdots, (z_1^{(N)},\cdots,z_l^{(N)})$.
\\
Let us set meromorphic function
$(f_m * f_n)(z_1^{(1)},\cdots, z_{m+n}^{(1)}|
\cdots|
z_1^{(N)},\cdots, z_{m+n}^{(N)})$
symmetric in each of variables
$(z_1^{(1)},\cdots, z_{m+n}^{(1)})$,$\cdots$
$(z_1^{(N)},\cdots, z_{m+n}^{(N)})$.
\begin{eqnarray}
&&(f_m * f_n)
(z_1^{(1)},\cdots,z_{m+n}^{(1)}|
\cdots|
z_1^{(N)},\cdots,z_{m+n}^{(N)})
\nonumber\\
&=&
\sum_{\sigma_1 \in S_{m+n}}
\sum_{\sigma_2 \in S_{m+n}}
\cdots
\sum_{\sigma_N \in S_{m+n}}\nonumber\\
&\times&f_m
(z_{\sigma_1(1)}^{(1)},\cdots,z_{\sigma_1(m)}^{(1)}|
\cdots|
z_{\sigma_N(1)}^{(N)},\cdots,
z_{\sigma_N(m)}^{(N)})
\nonumber\\
&\times&
f_n
(z_{\sigma_1(m+1)}^{(1)},\cdots,z_{\sigma_1(m+n)}^{(1)}|
\cdots|
z_{\sigma_N(m+1)}^{(N)},\cdots,
z_{\sigma_N(m+n)}^{(N)})
\nonumber\\
&\times&
\prod_{t=1}^N
\prod_{i=1}^m
\prod_{j=m+1}^{m+n}
\frac{\displaystyle
\left[
u_{\sigma_t(i)}^{(t)}-u_{\sigma_{t+1}(j)}^{(t+1)}+\frac{s}{N}
\right]_{r-1}
\left[
u_{\sigma_{t+1}(i)}^{(t+1)}-
u_{\sigma_{t}(j)}^{(t)}-1+\frac{s}{N}
\right]_{r-1}
}{
\displaystyle
\left[
u_{\sigma_t(i)}^{(t)}-u_{\sigma_{t}(j)}^{(t)}
\right]_{r-1}
\left[
u_{\sigma_{t}(j)}^{(t)}-
u_{\sigma_{t}(i)}^{(t)}+1\right]_{r-1}
}.\label{def:FO4}
\end{eqnarray}
Here meromorphic function
$f_l(z_1^{(1)},\cdots,z_l^{(1)}|
\cdots|
z_1^{(N)},\cdots,z_l^{(N)})$
is symmetric in each of variables
$(z_1^{(1)},\cdots,z_l^{(1)})$,
$\cdots, (z_1^{(N)},\cdots,z_l^{(N)})$.
\end{dfn}

We have a pair of infinitely many
commutative family of Feigin-Odesskii algebra.
Let us set theta function with parameters
$\nu_1,\cdots,\nu_N$ and
$\alpha$.
\begin{eqnarray}
&&\vartheta_{m,\alpha}(u_1^{(1)},\cdots,u_m^{(1)}|\cdots|
u_1^{(N)},\cdots,u_m^{(N)})=
\prod_{t=1}^N
\left[
\sum_{j=1}^m(u_j^{(t)}-u_j^{(t+1)})-\nu_t+\alpha
\right]_r.
\label{def:theta3}
\end{eqnarray}

\begin{prop}~~$\vartheta_{m,\vartheta}$ and 
$\vartheta_{n,\beta}$ commute
each other with respect to the product (\ref{def:FO3}).
\begin{eqnarray}
\vartheta_{m,\alpha} 
\circ 
\vartheta_{n,\beta}=
\vartheta_{n,\beta} \circ 
\vartheta_{m,\alpha}.
\end{eqnarray}
\end{prop}
Let us set theta function with parameters
$\nu_1,\cdots,\nu_N$ and
$\alpha$.
\begin{eqnarray}
&&\vartheta_{m,\alpha}^*(u_1^{(1)},\cdots,u_m^{(1)}|\cdots|
u_1^{(N)},\cdots,u_m^{(N)})=
\prod_{t=1}^N
\left[
\sum_{j=1}^m(u_j^{(t+1)}-u_j^{(t)})-\nu_t+\alpha
\right]_{r-1}.
\label{def:theta4}
\end{eqnarray}

\begin{prop}~~$\vartheta_{m,\alpha}$ and 
$\vartheta_{n,\beta}$ commute
each other with respect to the product (\ref{def:FO4}).
\begin{eqnarray}
\vartheta^*_{m,\alpha} * 
\vartheta^*_{n,\beta}=
\vartheta^*_{n,\beta} * 
\vartheta^*_{m,\alpha}.
\end{eqnarray}
\end{prop}
Proof of the above proposition is summarized in \cite{KS1}.

\subsection{Free field realization}

Let $\epsilon_j~(1\leqq j \leqq N)$ be an orthonormal basis
in ${\mathbb R}^N$ relative to the standard inner product $(\epsilon_i|\epsilon_j)
=\delta_{i,j}$.
Let us set $\bar{\epsilon}_j=\epsilon_j-\epsilon$
where $\epsilon=\frac{1}{N}\sum_{j=1}^N \epsilon_j$.
We identify $\epsilon_{j+N}=\epsilon_j$.
Let the weighted lattice $P=\sum_{j=1}^N{\mathbb Z}\bar{\epsilon}_j$.
Let us set $\alpha_j=\bar{\epsilon}_j-\bar{\epsilon}_{j+1}\in P$.
Let us introduce the bosons
$\beta_m^j~(m\in{\mathbb Z}_{\neq 0};1\leqq j \leqq N)$ by
\begin{eqnarray}
[\beta_m^i,\beta_n^j]=\left\{
\begin{array}{cc}
m\frac{[(r-1)m]}{[rm]}\frac{[(s-1)m]}{[sm]}\delta_{m+n,0},&
(i=j)\\
-m\frac{[(r-1)m]}{[rm]}\frac{[m]}{[sm]}x^{sm~{\rm sgn}(i-j)}
\delta_{m,n},& (i\neq j)
\end{array}
\right.
\end{eqnarray}
Let us set the commutation relations of $P_\lambda,Q_\mu~(\lambda,\mu \in P)$ by
\begin{eqnarray}
&&~[P_\lambda,iQ_\mu]=(\lambda|\mu).
\end{eqnarray}
We deal with
the bosonic Fock space
${\cal F}_{l,k}$,$(l,k \in P)$
generated by
$\beta_{-m}^i$,$(m>0, i=1,\cdots,N)$
over the vacuum vector $|l,k\rangle$.
\begin{eqnarray}
&&~\beta_m^i |l,k\rangle=0~(m>0,i=1,\cdots,N),\\
&&~P_\alpha|l,k\rangle=
\left(\alpha\left|
\sqrt{\frac{r}{(r-1)}}l-\sqrt{\frac{r-1}{r}}k\right.
\right)|l,k\rangle,\\
&&|l,k\rangle=
e^{\left(
i\sqrt{\frac{r}{(r-1)}}Q_l-i
\sqrt{\frac{r-1}{r}}Q_k
\right)}|0,0\rangle.
\end{eqnarray}

\begin{dfn}~~~We set the screening currents 
$F_j(z), (1\leqq j \leqq N)$ by
\begin{eqnarray}
F_j(z)&=&e^{i\sqrt{\frac{r-1}{r}}Q_{\alpha_j}}
(x^{(\frac{2s}{N}-1)j}z)^{\sqrt{\frac{r-1}{r}}P_{\alpha_j}+\frac{r-1}{r}}
\nonumber\\
&\times&
:\exp\left(\sum_{m\neq 0}\frac{1}{m}B_m^jz^{-m}\right):,~(1\leqq j\leqq N-1)\\
F_N(z)&=&e^{i\sqrt{\frac{r-1}{r}}Q_{\alpha_N}}
(x^{2s-N}z)^{\sqrt{\frac{r-1}{r}}P_{\bar{\epsilon}_N}+\frac{r-1}{2r}}
z^{-\sqrt{\frac{r-1}{r}}P_{\bar{\epsilon}_1}+\frac{r-1}{2r}}\nonumber\\
&\times&
:\exp\left(
\sum_{m\neq 0}\frac{1}{m}B_m^N
z^{-m}
\right):,
\end{eqnarray}
We set the screening currents $E_j(z), (1\leqq j \leqq N)$ by
\begin{eqnarray}
E_j(z)&=&e^{-i\sqrt{\frac{r}{r-1}}Q_{\alpha_j}}
(x^{(\frac{2s}{N}-1)j}z)
^{-\sqrt{\frac{r}{r-1}}P_{\alpha_j}+\frac{r}{r-1}}
\nonumber\\
&\times&
:\exp\left(-\sum_{m\neq 0}\frac{1}{m}
\frac{[rm]}{[(r-1)m]}
B_m^jz^{-m}\right):,~(1\leqq j\leqq N-1)\\
E_N(z)&=&e^{-i\sqrt{\frac{r}{r-1}}Q_{\alpha_N}}
(x^{2s-N}z)
^{-\sqrt{\frac{r}{r-1}}P_{\bar{\epsilon}_N}+\frac{r}{2(r-1)}}
z^{\sqrt{\frac{r}{r-1}}P_{\bar{\epsilon}_1}+\frac{r}{2(r-1)}}\nonumber\\
&\times&
:\exp\left(-
\sum_{m\neq 0}\frac{1}{m}
\frac{[rm]}{[(r-1)m]}B_m^N
z^{-m}
\right):.
\end{eqnarray}
Here we have set
\begin{eqnarray}
B_m^j&=&(\beta_m^j-\beta_m^{j+1})x^{-\frac{2s}{N}
jm},~(1\leqq j \leqq N-1),\\
B_m^N&=&(x^{-2sm}\beta_m^N-\beta_m^1).
\end{eqnarray}
\end{dfn}

\begin{prop}~~~
The currents $F_j(z),~(1\leqq j \leqq N;
N \geqq 3)$ satisfy
the following commutation relations.
\begin{eqnarray}
~\left[u_1-u_2-\frac{s}{N}\right]_r
F_j(z_1)F_{j+1}(z_2)
&=&\left[u_2-u_1+\frac{s}{N}-1\right]_r
F_{j+1}(z_2)F_{j}(z_1),~~(1\leqq j \leqq N),\nonumber\\\\
~[u_1-u_2]_r [u_1-u_2+1]_r
F_j(z_1)F_{j}(z_2)&=&[u_2-u_1]_r [u_2-u_1+1]_r
F_{j}(z_2)F_{j}(z_1),~~(1\leqq j \leqq N),
\nonumber\\
\\
F_i(z_1)F_j(z_2)&=&F_j(z_2)F_i(z_1),~~(|i-j|\geqq 2).
\end{eqnarray}
We read $F_{N+1}(z)=F_1(z)$. 
%%%%%%%%%%%%%%%%%%%%%%%%%%%%%%%%%%%%%%%%%%%%
%%%%%%%%%%%%%%%%%%%%%%%%%%%%%%%%%%%%%%%%%%%%
The currents $E_j(z),~(1\leqq j \leqq N;
N \geqq 3)$ satisfy
the following commutation relations.
\begin{eqnarray}
\left[u_1-u_2+1-\frac{s}{N}\right]_{r-1}
E_j(z_1)E_{j+1}(z_2)
&=&
\left[u_2-u_1+\frac{s}{N}\right]_{r-1}
E_{j+1}(z_2)E_{j}(z_1),~~(1\leqq j \leqq N),\nonumber\\
\\
~[u_1-u_2]_{r-1}[u_1-u_2-1]_{r-1}
E_j(z_1)E_{j}(z_2)&=&[u_2-u_1]_{r-1}[u_2-u_1-1]_{r-1}
E_{j}(z_2)E_{j}(z_1),~~(1\leqq j \leqq N),\nonumber\\
\\
E_i(z_1)E_j(z_2)&=&E_j(z_2)E_i(z_1),~~(|i-j|\geqq 2).
\end{eqnarray}
We read $E_{N+1}(z)=E_1(z)$. 
\end{prop}

\begin{prop}~~
The screening currents
$E_j(z), F_j(z)$, $(1\leqq j \leqq N; N\geqq 3)$
satisfy the following relation.
\begin{eqnarray}
[E_i(z_1),F_j(z_2)]
&=&\frac{\delta_{i,j}}{x-x^{-1}}
(\delta(xz_2/z_1){H}_j(x^{r}z_2)
-\delta(xz_1/z_2){H}_j(x^{-r}z_2)),
(1\leqq i,j \leqq N).\nonumber\\
\end{eqnarray}
Here we have set
\begin{eqnarray}
{H}_j(z)&=&x^{(1-\frac{2s}{N})2j}
e^{-\frac{i}{\sqrt{r (r-1)}}Q_{\alpha_j}}
(x^{(\frac{2s}{N}-1)j}z)^{-\frac{1}{
\sqrt{r (r-1)}}P_{\alpha_j}+\frac{1}{r (r-1)}}\nonumber\\
&\times&:\exp\left(
-\sum_{m\neq 0}\frac{1}{m}\frac{[m]}{
[(r-1)m]}B_m^j z^{-m}\right):
,~~(1\leqq j \leqq N-1),\\
{H}_N(z)&=&x^{2(N-2s)}
e^{-\frac{i}{\sqrt{r (r-1)}}Q_{\alpha_N}}
(x^{2s-N}z)^{-\frac{1}{\sqrt{r (r-1)}}P_{{\bar{\epsilon}}_N}
+\frac{1}{2 r (r-1)}}
z^{-\frac{1}{\sqrt{r (r-1)}}P_{{\bar{\epsilon}}_1}
+\frac{1}{2 r (r-1)}}\nonumber\\
&\times&
:\exp\left(
-\sum_{m\neq 0}\frac{1}{m}\frac{[m]}{
[(r-1)m]}B_m^N z^{-m}\right):.
\end{eqnarray}
\end{prop}

\begin{dfn}~~
Let us set the functional ${\cal G}$ by
\begin{eqnarray}
{\cal G}(f_m)
&=&\oint \cdots \oint \prod_{t=1}^N \prod_{j=1}^m
\frac{dz_j^{(t)}}{2\pi i z_j^{(t)}}
F_1(z_1^{(1)}) \cdots F_1(z_m^{(1)})\cdots
F_N(z_1^{(N)}) \cdots F_N(z_m^{(N)})\nonumber\\
&\times&
\frac{\displaystyle
\prod_{t=1}^N \prod_{1\leqq i<j \leqq m}
\left[u_i^{(t)}-u_j^{(t)}\right]_r
\left[u_j^{(t)}-u_i^{(t)}-1\right]_r}{
\displaystyle
\prod_{t=1}^{N-1}\prod_{i,j=1}^m
\left[u_i^{(t)}-u_j^{(t+1)}+1-\frac{s}{N}\right]_r
\prod_{i,j=1}^m
\left[u_i^{(1)}-u_j^{(N)}+\frac{s}{N}\right]_r}\nonumber\\
&\times&
f_m(z_1^{(1)},\cdots,z_m^{(1)}|\cdots|
z_1^{(N)},\cdots,z_m^{(N)}).
\label{def:functional3}
\end{eqnarray}
We take the integration contours to be simple closed curves
around the origin satisfying
\begin{eqnarray}
&&|x^{\frac{2s}{N}}z_j^{(t+1)}|<
|z_i^{(t)}|<
|x^{-2+\frac{2s}{N}}z_j^{(t+1)}|,~
(1\leqq t \leqq N-1,1\leqq i,j \leqq m),\nonumber
\\
&&|x^{2^-\frac{2s}{N}}z_j^{(1)}|<|z_i^{(N)}|<
|x^{-\frac{2s}{N}}z_j^{(1)}|,~(1\leqq i,j \leqq m).\nonumber
\end{eqnarray}
Let us set the functional
${\cal G}^*$ by followings.
\begin{eqnarray}
{\cal G}^*(f_m)
&=&\oint \cdots \oint \prod_{t=1}^N \prod_{j=1}^m
\frac{dz_j^{(t)}}{2\pi i z_j^{(t)}}
E_1(z_1^{(1)}) \cdots E_1(z_m^{(1)})\cdots
E_N(z_1^{(N)}) \cdots E_N(z_m^{(N)})\nonumber\\
&\times&
\frac{\displaystyle
\prod_{t=1}^N \prod_{1\leqq i<j \leqq m}
\left[u_i^{(t)}-u_j^{(t)}\right]_{r-1}
\left[u_j^{(t)}-u_i^{(t)}-1\right]_{r-1}}{
\displaystyle
\prod_{t=1}^{N-1}\prod_{i,j=1}^m
\left[u_i^{(t)}-u_j^{(t+1)}-1+\frac{s}{N}\right]_{r-1}
\prod_{i,j=1}^m
\left[u_i^{(1)}-u_j^{(N)}-\frac{s}{N}\right]_{r-1}}\nonumber\\
&\times&
f_m(z_1^{(1)},\cdots,z_m^{(1)}|\cdots|
z_1^{(N)},\cdots,z_m^{(N)}).
\label{def:functional4}
\end{eqnarray}
We take the integration contours to be simple closed curves
around the origin satisfying
\begin{eqnarray}
&&|x^{\frac{2s}{N}}z_j^{(t+1)}|<
|z_i^{(t)}|<
|x^{-2+\frac{2s}{N}}z_j^{(t+1)}|,~
(1\leqq t \leqq N-1,1\leqq i,j \leqq m),\nonumber
\\
&&|x^{2^-\frac{2s}{N}}z_j^{(1)}|<|z_i^{(N)}|<
|x^{-\frac{2s}{N}}z_j^{(1)}|,~(1\leqq i,j \leqq m).\nonumber
\end{eqnarray}
\end{dfn}

\begin{prop}~~
When the functions 
$f_l(z_1^{(1)},\cdots,z_l^{(1)}|\cdots |
z_1^{(N)},\cdots,z_l^{(N)})$
is meromorphic function symmetric in each of varibles
$(z_1^{(t)},\cdots,z_l^{(t)})$, $(1\leqq t \leqq N)$, and
don't have poles at the origin $z_j^{(t)}=0$, 
$(1\leqq t \leqq N, 1\leqq j \leqq l)$,
the functionals ${\cal G}$, ${\cal G}^*$ satisfy
\begin{eqnarray}
&&{\cal G}(f_m){\cal G}(f_n)=
{\cal G}(f_m \circ f_n),\\
&&{\cal G}^*(f_m){\cal G^*}(f_n)=
{\cal G}^*(f_m * f_n).
\end{eqnarray}
\end{prop}
Symmetrizing with respect to the integration variables
$(z_1^{(t)},\cdots,z_{m+n}^{(t)})$
product ${\cal G}(f_m){\cal G}(f_n)$, 
we have the above proposition.

In what follows we set parameters in the theta function 
$\vartheta_{m,\alpha}$,
$\vartheta_{m,\alpha}^*$ ;
$\nu_t=\sqrt{r(r-1)}P_{\bar{\epsilon}_{t+1}}$, 
$(1\leqq t \leqq N)$,
$\alpha=\sum_{t=1}^N \alpha_t P_{\bar{\epsilon}_t}$,
$(\alpha_t \in {\mathbb C})$.
Because the relation $\sum_{t=1}^N P_{\bar{\epsilon}_t}=0$,
$\vartheta_{m,\alpha}$,
$\vartheta_{m,\alpha}^*$
have $(N-1)$ independent parameters.

\begin{thm}~~For $r>1$ we have
\begin{eqnarray}
&&[{\cal G}(\vartheta_{m,\alpha}),
{\cal G}(\vartheta_{n,\beta})]=0,~~
(m,n \in {\mathbb N}),\\
&&[{\cal G}^*(\vartheta_{m,\alpha}^*),
{\cal G}^*(\vartheta^*_{n,\beta})]=0,~~
(m,n \in {\mathbb N}).
\end{eqnarray}
\end{thm}

\begin{thm}~~For $0<r<1$ we have
\begin{eqnarray}
[{\cal G}(\vartheta_{m,\alpha}),
{\cal G}^*(\vartheta^*_{n,\beta})]=0,~~
(m,n \in {\mathbb N}).
\end{eqnarray}
\end{thm}
Definition of ${\cal G}^*(\vartheta^*_{m,\alpha})$
for $0<r<1$ is given as the same manner as 
(\ref{def:functional2}). See detailds in \cite{KS1}.
We have constructed infinitely many
commutative operators
${\cal G}(\vartheta_{m,\alpha})$, 
${\cal G}^*(\vartheta_{m,\alpha}^*)$,
$(m \in {\mathbb N})$ 
acting on the bosonic Fock space,
which is regarded as the free field realization
of commutative family of
Feigin-Odesskii algebra (\ref{def:FO3}) and (\ref{def:FO4}).

\section{Level $k$ generalization of $U_{q,p}(\widehat{sl_2})$}

In this section we consider level $k$ generaliztion
of section 2.
Main contribution is construction of
free field realization for
one parameter $s$ deformation of
Level $k$ elliptic albegra 
$U_{q,p}(\widehat{sl_2})$.

\subsection{Feigin-Odesskii algebra}

Let us set parameters $r,k \in {\mathbb R}$ 
such that $r>0, r-k>0$.
It's not difficult to give Level $k$ generalization of
Feigin-Odesskii algebra:
$f \circ g$ and $f * g$.

\begin{dfn}~~
Let us set the symmetric function 
$(f_m \circ f_n)(z_1,\cdots,z_{m+n}|w_1,\cdots,w_{m+n})$ 
by the same relation (\ref{def:FO1}).

Let us set the symmetric function
$(f_m * f_n)(z_1 \cdots z_{m+n}|w_1 \cdots w_{m+n})$ by 
modification of (\ref{def:FO2}).
\begin{eqnarray}
&&(f_m * f_n)(z_1,\cdots,z_{m+n}|
w_1,\cdots,w_{m+n})\nonumber\\
&=&
\frac{1}{((m+n)!)^2}
\sum_{\sigma \in S_{m+n}}
\sum_{\tau \in S_{m+n}}
f_m(z_{\sigma(1)},\cdots,z_{\sigma(m)}|
w_{\tau(1)},\cdots,w_{\tau(m)})\nonumber\\
&&\times
f_n(z_{\sigma(m+1)},\cdots,z_{\sigma(m+n)}|
w_{\tau(m+1)},\cdots,w_{\tau(m+n)})\nonumber
\\
&&\times
\prod_{i=1}^m \prod_{j=m+1}^{m+n}
\frac
{\displaystyle
\left[v_{\tau(i)}-u_{\sigma(j)}-\frac{s}{2}\right]_{r-k}
\left[u_{\sigma(i)}-v_{\tau(j)}-\frac{s}{2}\right]_{r-k}
}
{
\displaystyle
[u_{\sigma(i)}-u_{\sigma(j)}]_{r-k}
[u_{\sigma(j)}-u_{\sigma(i)}+1]_{r-k}
}\nonumber\\
&\times&
\prod_{i=1}^m \prod_{j=m+1}^{m+n}
\frac
{\displaystyle
\left[u_{\sigma(j)}-v_{\tau(i)}-\frac{s}{2}+1\right]_{r-k}
\left[v_{\tau(j)}-u_{\sigma(i)}-\frac{s}{2}+1\right]_{r-k}}
{
\displaystyle
[v_{\tau(j)}-v_{\tau(i)}+1]_{r-k}
[v_{\tau(j)}-v_{\tau(i)}+1]_{r-k}}.\label{def:FO5}
\end{eqnarray}
Here $f_l(z_1,\cdots,z_l|w_1,\cdots,w_l)$
are meromorphic function symmetric in each of varibles
$(z_1,\cdots,z_l)$ and $(w_1,\cdots,w_l)$.
\end{dfn}
We have infinitely many
commutative solutions 
$\vartheta_{m,\alpha}$ and
$\vartheta_{m,\alpha}^*$
with respect with product $f \circ g$
and $f * g$.
The solutions $\vartheta_{m,\alpha}(z_1,\cdots,z_m)$ 
for product $\circ$ is
given as the same as (\ref{def:theta1}).
Let us set the theta function
$\vartheta_{m,\alpha}^*$ 
with parmeters $\alpha, \nu$.
\begin{eqnarray}
&&\vartheta_m(z_1,\cdots,z_m|w_1,\cdots,w_m)=
\left[\sum_{j=1}^m(v_j-u_j)-\nu+\alpha \right]_{r-k}
\left[\sum_{j=1}^m(u_j-v_j)-\alpha \right]_{r-k}.
\end{eqnarray}

\begin{prop}~~$\vartheta_{m,alpha}$ and $\vartheta_{n,\beta}$
commute with respect to the product (\ref{def:FO5}).
\begin{eqnarray}
\vartheta_{m,\alpha}*{\vartheta}_{n,\beta}=
\vartheta_{n,\beta}*{\vartheta}_{m,\alpha}.
\end{eqnarray}
\end{prop}

\subsection{Free field realization}

In this section we give
one parameter deformation of
Wakimoto realization of elliptic algebra 
$U_{q,p}(\widehat{sl_2})$ \cite{Matsuo, Konno}.
Let us set deformation parameter $0<s<2$.
Let us set
the bosons $\alpha_m^j,\widetilde{\alpha}_m^j, (j=1,2; m\in {\mathbb Z}_{\neq 0})$, 
\begin{eqnarray}
~[\alpha_m^j,\alpha_n^j]&=&-\frac{1}{m}
\frac{[2m][rm]}{[km]
[(r-k)m]}\delta_{m+n,0},~~(j=1,2),\\
~[\alpha_m^1,\alpha_{n}^2]&=&
\frac{1}{m}
\left(
\frac{x^{(-r+k)m}([sm]-[(s-2)m])}{[(r-k)m]}
+
\frac{x^{km}([sm]+[(s-2)m])}{[km]}
\right)\delta_{m+n,0},\\
~[\widetilde{\alpha}_m^j,
\widetilde{\alpha}_n^j]&=&-\frac{1}{m}
\frac{[2m][(r-k)m]}{[km]
[rm]}\delta_{m+n,0},~~(j=1,2),\\
%%%%%%%%%%%%%%%%%%%%%%%%%%%%%%%%%%%%%%%%%%%
%~[\widetilde{\alpha}_m^1,
%\widetilde{\alpha}_n^2]
%&=&\frac{1}{m}\left(
%\frac{[sm][(r-k)m]
%}{[km][rm]}+
%\frac{[(s-2)m][(r-k)m]_+
%}{[km][rm]}
%\right)\delta_{m+n,0},\\
~[\widetilde{\alpha}_m^1,
\widetilde{\alpha}_{n}^2]&=&
\frac{1}{m}
\left(
\frac{x^{rm}(-[sm]+[(s-2)m])}{[rm]}
+
\frac{x^{km}
([sm]+[(s-2)m])}{[km]}
\right)\delta_{m+n,0},\\
~[\alpha_m^j,\widetilde{\alpha}_n^j]&=&
-\frac{1}{m}
\frac{[2m]}{[km]}\delta_{m+n,0},~~(j=1,2),\\
~[\alpha_m^1,\widetilde{\alpha}_n^2]&=&
\frac{1}{m}
\frac{[sm]+[(s-2)m]
}{[km]}\delta_{m+n,0},\\
~[\widetilde{\alpha}_m^1,{\alpha}_n^2]&=&
\frac{1}{m}
\frac{[sm]+[(s-2)m]
}{[km]}\delta_{m+n,0}.
\end{eqnarray}
We set
the bosons $\beta_m^j, \gamma_m^j$, 
$(j=1,2; m \in {\mathbb Z}_{\neq 0})$,
\begin{eqnarray}
~[\beta_m^j,\beta_n^j]&=&
\frac{[2m][(k+2)m]}{m}
\delta_{m+n,0},~(j=1,2),\\
~[\beta_m^1,\beta_n^2]&=&
-\frac{[(k+2)m]
([sm]+[(s-2)m])
}{m}
\delta_{m+n,0},\\
~[\gamma_m^j,\gamma_n^j]&=&
\frac{1}{m}
\frac{[2m]}{[km]}\delta_{m+n,0},~~(j=1,2),\\
~[\gamma_m^1,\gamma_n^2]&=&
-\frac{1}{m}
\frac{[sm]+[(s-2)m]}
{[km]}
\delta_{m+n,0}.
\end{eqnarray}
We set the zero-mode operators $P_0, Q_0$,
$h,\alpha$ and $h_0,h_1,h_2,\alpha_0,\alpha_1,\alpha_2$,
\begin{eqnarray}
&&~[P_0,iQ_0]=1,~[h,\alpha]=2,\\
&&~~[h_0,\alpha_0]=[h_1,\alpha_2]=[h_2,\alpha_1]=(2-s),~~
[h_1,\alpha_1]=[h_2,\alpha_2]=0.
\end{eqnarray}
We set the Fock space ${\cal F}_{K,L}$, $(K,L \in {\mathbb Z})$.
\begin{eqnarray}
&&{\cal F}_{K,L}=
\bigoplus_{n,n_0,n_1,n_2 \in {\mathbb Z}}
{\mathbb C}[\alpha_{-m}^j,\widetilde{\alpha}_{-m}^j,
\beta_{-m}^j, \gamma_{-m}^j,(j=1,2; m \in {\mathbb N}_{\neq 0})
]\otimes 
|K,L\rangle_{n,n_0,n_1,n_2},\nonumber\\
\\
&&|K,L \rangle_{n,n_0,n_1,n_2}=
e^{\left(L\sqrt{\frac{r}{2(r-k)}}-K
\sqrt{\frac{r-k}{2r}}\right)iQ}\otimes
e^{n \alpha}\otimes e^{n_0 \alpha_0}
\otimes e^{n_1 \alpha_1}\otimes e^{n_2 \alpha_2}.
\end{eqnarray}
Upon specialization $s\to 2$,
simplification occures.
\begin{eqnarray}
&&\alpha_m^2=-\alpha_m^1,~~
\widetilde{\alpha}_m^1=\frac{[(r-k)m]}{[rm]}\alpha_m^1,~~
\widetilde{\alpha}_m^2=-\frac{[(r-k)m]}{[rm]}\alpha_m^1,\\
&&\beta_m^2=-\beta_m^1,~~
\gamma_m^2=-\gamma_m^1,~~~
h_0=h_1=h_2=\alpha_0=\alpha_1=\alpha_2=0.
\end{eqnarray}
The bosons $\alpha_m^1,\beta_m^1,\gamma_m^1$ are the same bosons
which were introduced to construct the elliptic current associated with
the elliptic algebra $U_{q,p}(\widehat{sl_2})$
\cite{Matsuo, Konno, JKOS}.
In order to construct
infinitly many commutative operators,
we introduce one prameter $s$ deformation of the bosons in 
\cite{Matsuo, Konno, JKOS}.
We introduce the operators 
$C_j(z), C_j^\dagger(z)$, $(j=1,2)$ acting on
the Fock space ${\cal F}_{J,K}$.
\begin{eqnarray}
C_1(z)&=&e^{-\sqrt{\frac{2r}{k(r-k)}}iQ_0}
e^{-\sqrt{\frac{2r}{k(r-k)}}P_0{\rm log}z}
:\exp\left(-\sum_{m \neq 0}\alpha_{m}^1z^{-m}
\right):,
\\
C_2(z)&=&
e^{\sqrt{\frac{2r}{k(r-k)}}iQ_0}
e^{\sqrt{\frac{2r}{k(r-k)}}P_0{\rm log}z}
:\exp\left(-\sum_{m \neq 0}\alpha_{m}^2z^{-m}\right):,
\\
C_1^\dagger(z)&=&
e^{\sqrt{\frac{2(r-k)}{kr}}iQ_0}
e^{\sqrt{\frac{2(r-k)}{kr}}P_0{\rm log}z}
:\exp\left(\sum_{m \neq 0}
\widetilde{\alpha}_{m}^1z^{-m}\right):,
\\
C_2^\dagger(z)&=&
e^{-\sqrt{\frac{2(r-k)}{kr}}iQ_0}
e^{-\sqrt{\frac{2(r-k)}{kr}}P_0{\rm log}z}
:\exp\left(\sum_{m \neq 0}
\widetilde{\alpha}_{m}^2z^{-m}\right):.
\end{eqnarray}
We set the operators $\widetilde{\Psi}_{j,I}(z), 
\widetilde{\Psi}_{j,II}(z), 
\widetilde{\Psi}_{j,I}^\dagger(z),
\widetilde{\Psi}_{j,II}^\dagger(z)
$, $(j=1,2)$ acting on
the Fock space ${\cal F}_{J,K}$.
\begin{eqnarray}
\widetilde{\Psi}_{j,I}(z)
&=&%x^{\frac{1}{2}(-P_1-P_2)}
\exp\left(-(x-x^{-1})
\sum_{m>0}\frac{x^{\frac{km}{2}}}{[m]_+}\beta_{m}^jz^{-m}\right)
\\
&\times&
\exp
\left(-
\sum_{m>0}x^{-\frac{km}{2}}\gamma_{-m}^jz^{m}\right)
\exp
\left(-
\sum_{m>0}x^{\frac{km}{2}}\frac{[(k+1)m]_+}{[m]_+}
\gamma_{m}^jz^{-m}\right),~(j=1,2),\nonumber
\\
\widetilde{\Psi}_{j,II}(z)
&=&%x^{\frac{1}{2}(P_1+P_2)}
\exp\left((x-x^{-1})
\sum_{m>0}
\frac{x^{\frac{km}{2}}}{[m]_+}\beta_{-m}^jz^{m}\right)
\\
&\times&
\exp
\left(-
\sum_{m>0}x^{\frac{km}{2}}
\frac{[(k+1)m]_+}{[m]_+}
\gamma_{-m}^j z^{m}\right)
\exp
\left(-
\sum_{m>0}x^{-\frac{km}{2}}
\gamma_{m}^j
z^{-m}\right),~(j=1,2),\nonumber\\
\widetilde{\Psi}_{j,I}^\dagger(z)
&=&%x^{\frac{1}{2}(P_1-P_2)}
\exp\left((x-x^{-1})
\sum_{m>0}
\frac{x^{-\frac{km}{2}}}{[m]_+}\beta_{m}^j
z^{-m}\right)
\\
&\times&
\exp
\left(
\sum_{m>0}x^{\frac{km}{2}}\gamma_{-m}^j
z^{m}\right)
\exp
\left(
\sum_{m>0}
x^{-\frac{km}{2}}\frac{[(k+1)m]_+}{[m]_+}
\gamma_{m}^jz^{-m}\right),~(j=1,2),\nonumber\\
\widetilde{\Psi}_{j,II}^\dagger(z)
&=&%x^{\frac{1}{2}(-P_1+P_2)}
\exp\left(-(x-x^{-1})
\sum_{m>0}
\frac{x^{-\frac{km}{2}}}{[m]_+}
\beta_{-m}^j
z^{m}\right)
\\
&\times&
\exp
\left(
\sum_{m>0}x^{-\frac{km}{2}}
\frac{[(k+1)m]_+}{[m]_+}
\gamma_{-m}^j
z^{m}\right)
\exp
\left(
\sum_{m>0}x^{\frac{km}{2}}
\gamma_{m}^j
z^{-m}\right),~(j=1,2).\nonumber
\end{eqnarray}
We set the operators $\Psi_{j,I}(z), 
\Psi_{j,II}(z), 
\Psi_{j,I}^\dagger(z),
\Psi_{j,II}^\dagger(z)
$, $(j=1,2)$ acting on
the Fock space ${\cal F}_{J,K}$.
\begin{eqnarray}
\Psi_{1,I}(z)&=&
\widetilde{\Psi}_{1,I}(z)
e^{\alpha+\alpha_0+\alpha_1}
x^{\frac{h}{2}+h_0+h_1}z^{-\frac{h}{k}},\\
\Psi_{1,II}(z)&=&
\widetilde{\Psi}_{1,II}(z)
e^{\alpha+\alpha_0+\alpha_1}
x^{-\frac{h}{2}+h_0-h_1}z^{-\frac{h}{k}},\\
\Psi_{2,I}(z)&=&
\widetilde{\Psi}_{2,I}(z)
e^{-\alpha-\alpha_0+\alpha_2}
x^{-\frac{h}{2}+h_0+h_2}z^{\frac{h}{k}},\\ 
\Psi_{2,II}(z)&=&
\widetilde{\Psi}_{2,II}(z)
e^{-\alpha-\alpha_0+\alpha_2}
x^{\frac{h}{2}+h_0-h_2}z^{\frac{h}{k}},\\
%%%%%%%%%%%%%%%%%%%%%%%%%%%%%%%%%%%%%%%%%%%%%%%%%%
%%%%%%%%%%%%%%%%%%%%%%%%%%%%%%%%%%%%%%%%%%%%%%%%%%
\Psi_{1,I}^\dagger(z)&=&
\widetilde{\Psi}_{1,I}^\dagger(z)
e^{-\alpha-\alpha_0+\alpha_1}
x^{\frac{h}{2}-h_0-h_1}z^{\frac{h}{k}},\\
\Psi_{1,II}^\dagger(z)&=&
\widetilde{\Psi}_{1,II}^\dagger(z)
e^{-\alpha-\alpha_0+\alpha_1}
x^{-\frac{h}{2}-h_0+h_1}z^{\frac{h}{k}},\\
\Psi_{2,I}^\dagger(z)&=&
\widetilde{\Psi}_{2,I}^\dagger(z)
e^{\alpha+\alpha_0+\alpha_2}
x^{-\frac{h}{2}-h_0-h_2}z^{-\frac{h}{k}},\\ 
\Psi_{2,II}^\dagger(z)&=&
\widetilde{\Psi}_{2,II}^\dagger(z)
e^{\alpha+\alpha_0+\alpha_2}
x^{\frac{h}{2}-h_0+h_2}z^{-\frac{h}{k}}.
\end{eqnarray}

\begin{dfn}~~We set the operators 
$E_j(z), F_j(z)$, $(j=1,2)$,
which can be regarded as one parameter
deformation of the level $k$ elliptic currents 
associated with the elliptic algebra $U_{q,p}(\widehat{sl_2})$
\cite{Konno, JKOS}.
\begin{eqnarray}
E_j(z)=C_j(z)\Psi_j(z),~~
F_j(z)=C_j^\dagger(z)\Psi_j^\dagger(z),~~(j=1,2),
\end{eqnarray}
where we have set
\begin{eqnarray}
\Psi_j(z)=\frac{1}{x-x^{-1}}(\Psi_{j,I}(z)-\Psi_{j,II}(z)),~~
\Psi_j^\dagger(z)=\frac{-1}{x-x^{-1}}(\Psi_{j,I}^\dagger(z)-
\Psi_{j,II}^\dagger(z)),~~
(j=1,2).
\end{eqnarray}
\end{dfn}

\begin{prop}~~The elliptic currents $E_j(z)$, $(j=1,2)$ satisfy
the following commutation relations.
\begin{eqnarray}
~&&[u_1-u_2]_{r-k}[u_1-u_2-1]_{r-k}E_j(z_1)E_j(z_2)\nonumber
\\&=&[u_2-u_1]_{r-k}[u_2-u_1-1]_{r-k}E_j(z_2)E_j(z_1),~(j=1,2),
\\
~&&\left[u_1-u_2+\frac{s}{2}\right]_{r-k}
\left[u_1-u_2-\frac{s}{2}+1\right]_{r-k}
E_1(z_1)E_2(z_2)\nonumber\\
&=&
\left[u_2-u_1+\frac{s}{2}\right]_{r-k}
\left[u_2-u_1-\frac{s}{2}+1\right]_{r-k}
E_2(z_2)E_1(z_1).
\end{eqnarray}
The elliptic currents $F_j(z)$, $(j=1,2)$ satisfy
the following commutation relations.
%%%%%%%%%%%%%%%%%%%%%%%%%%%%%%%%%%%%%%%%%%%%%%%%%%%
\begin{eqnarray}
~&&[u_1-u_2]_r[u_1-u_2+1]_rF_j(z_1)F_j(z_2)\nonumber\\
&=&[u_2-u_1]_r[u_2-u_1+1]_rF_j(z_2)F_j(z_1),~(j=1,2),\\
~&&\left[
u_1-u_2-\frac{s}{2}
\right]_r\left[u_1-u_2+\frac{s}{2}-1\right]_r
F_1(z_1)F_2(z_2)\nonumber\\
&=&
\left[u_2-u_1-\frac{s}{2}\right]_r
\left[u_2-u_1+\frac{s}{2}-1\right]_r
F_2(z_2)F_1(z_1).
\end{eqnarray}
%%%%%%%%%%%%%%%%%%%%%%%%%%%%%%%%%%%%%%%%%%%%%%%%%%
The currents $E_j(z)$ and $F_j(z)$ satisfy
\begin{eqnarray}
~[E_j(z_1),F_j(z_2)]&=&\frac{x^{(-1)^j (s-2)}}{x-x^{-1}}
\left(:C_j(z_1)C_j^\dagger(z_2)
\Psi_{j,I}(z_1)\Psi_{j,I}^\dagger(z_2):
\delta\left(\frac{x^kz_2}{z_1}\right)\right.
\\
&-&
\left.
:C_j(z_1)C_j^\dagger(z_2)
\Psi_{j,II}(z_1)\Psi_{j,II}^\dagger(z_2):
\delta\left(\frac{x^{-k}z_2}{z_1}\right)\right),~~
(j=1,2).\nonumber
\end{eqnarray}
Here we have
used the delta-function $\delta(z)=\sum_{n \in {\mathbb Z}}z^n$.
\end{prop}

The definition
of the functional ${\cal G}(f_m)$
is given as the same as (\ref{def:FO1}).

\begin{dfn}~~
Let us set the functional
${\cal G}^*$ by followings.
\begin{eqnarray}
{\cal G}^*(f_m)
&=&
\oint \prod_{j=1}^m\frac{dz_j}{2\pi i z_j}
\oint \prod_{j=1}^m\frac{dw_j}{2\pi i w_j}
E_1(z_1)\cdots E_1(z_m)E_2(w_1)\cdots E_2(w_m)
\label{def:functional2}\\
&\times&
\frac{\displaystyle
\prod_{1\leqq j<k \leqq m}
[u_i-u_j]_{r-1}[u_j-u_i+1]_{r-k}
[v_i-v_j]_{r-1}[v_j-v_i+1]_{r-k}}{
\displaystyle
\prod_{i=1}^m
\prod_{j=1}^m
\left[u_i-v_j-\frac{s}{2}\right]_{r-k}
\left[v_j-u_i-\frac{s}{2}+1\right]_{r-k}
}f_m(z_1,\cdots,z_m|w_1,\cdots,w_m).\nonumber
\end{eqnarray}
We take the integration contours to be simple closed curves
around the origin satisfying
\begin{eqnarray}
|z_j^{(t)}|=1,~~(t=1,2; j=1,2,\cdots,m).\nonumber
\end{eqnarray}
\end{dfn}

\begin{prop}~~
When the function $f_l(z_1,\cdots,z_l|w_1,\cdots,w_l)$
are meromorphic function symmetric in each of varibles
$(z_1,\cdots,z_l)$, $(w_1,\cdots,w_l)$, and
don't have poles at the origin $z_j=0$, $w_j=0$,
the functionals ${\cal G}$, ${\cal G}^*$ satisfy
\begin{eqnarray}
&&{\cal G}(f_m){\cal G}(f_n)=
{\cal G}(f_m \circ f_n),\\
&&{\cal G}^*(f_m){\cal G^*}(f_n)=
{\cal G}^*(f_m*f_n).
\end{eqnarray}
\end{prop}
In what follows we set parameters 
$\nu=\sqrt{\frac{2r(r-k)}{k}}P_0+\frac{r-k}{r}h$ 
in theta function $\vartheta_{m,\alpha},
\vartheta_{m,\alpha}^*$.

\begin{thm}~~For $r>0$ and $r-k>0$, we have
\begin{eqnarray}
&&[{\cal G}(\vartheta_{m,\alpha}),
{\cal G}(\vartheta_{n,\beta})]=0,~~
(m,n \in {\mathbb N}),\\
&&[{\cal G}^*(\vartheta_{m,\alpha}^*),{\cal G}^*
(\vartheta^*_{n,\beta})]=0,~~
(m,n \in {\mathbb N}).
\end{eqnarray}
\end{thm}
We have constructed infinitely many
commutative operators
${\cal G}(\vartheta_{m,\alpha})$, 
${\cal G}^*(\vartheta_{m,\alpha}^*)$,
$(m \in {\mathbb N})$ 
acting on the bosonic Fock space,
which is regarded as the free field realization
of commutative family of
Feigin-Odesskii algebra (\ref{def:FO1}) and (\ref{def:FO5}).

%%%%%%%%%%%%%%%%%%%%%%%%%%%%%%%%%%%%%%%%%%%%%%%%

\section{Level $k$ generalization of 
$U_{q,p}(\widehat{sl_N})$}

In this section we 
report some results for Level $k$ generalization of
section 3, which are now in progress.
Main result is free field realization of 
Level $k$ elliptic algebra $U_{q,p}(\widehat{sl_N})$.

%%%%%%%%%%%%%%%%%%%%%%%%%%%%%%%%%%%%
%%%%%%%%%%%%%%%%%%%%%%%%%%%%%%%%%%%%

\subsection{Feigin-Odesskii algebra}

We introduce a pair of Feigin-Odesskii algebra.

\begin{dfn}~~~
Let us set meromorphic function
$(f_m * f_n)(z_1^{(1)},\cdots, z_{m+n}^{(1)}|
\cdots|
z_1^{(N)},\cdots, z_{m+n}^{(N)})$
symmetric in each of variables
$(z_1^{(1)},\cdots, z_{m+n}^{(1)})$,$\cdots$
$(z_1^{(N)},\cdots, z_{m+n}^{(N)})$.
\begin{eqnarray}
&&(f_m * f_n)
(z_1^{(1)},\cdots,z_{m+n}^{(1)}|
\cdots|
z_1^{(N)},\cdots,z_{m+n}^{(N)})
\nonumber\\
&=&
\sum_{\sigma_1 \in S_{m+n}}
\sum_{\sigma_2 \in S_{m+n}}
\cdots
\sum_{\sigma_N \in S_{m+n}}\nonumber\\
&\times&f_m
(z_{\sigma_1(1)}^{(1)},\cdots,z_{\sigma_1(m)}^{(1)}|
\cdots|
z_{\sigma_N(1)}^{(N)},\cdots,
z_{\sigma_N(m)}^{(N)})
\nonumber\\
&\times&
f_n
(z_{\sigma_1(m+1)}^{(1)},\cdots,z_{\sigma_1(m+n)}^{(1)}|
\cdots|
z_{\sigma_N(m+1)}^{(N)},\cdots,
z_{\sigma_N(m+n)}^{(N)})
\nonumber\\
&\times&
\prod_{t=1}^N
\prod_{i=1}^m
\prod_{j=m+1}^{m+n}
\frac{\displaystyle
\left[
u_{\sigma_t(i)}^{(t)}-u_{\sigma_{t+1}(j)}^{(t+1)}+\frac{s}{N}
\right]_{r-k}
\left[
u_{\sigma_{t+1}(i)}^{(t+1)}-
u_{\sigma_{t}(j)}^{(t)}-1+\frac{s}{N}
\right]_{r-k}
}{
\displaystyle
\left[
u_{\sigma_t(i)}^{(t)}-u_{\sigma_{t}(j)}^{(t)}
\right]_{r-k}
\left[
u_{\sigma_{t}(j)}^{(t)}-
u_{\sigma_{t}(i)}^{(t)}+1\right]_{r-k}
}.\label{def:FO6}
\end{eqnarray}
Here meromorphic function
$f_l(z_1^{(1)},\cdots,z_l^{(1)}|
\cdots|
z_1^{(N)},\cdots,z_l^{(N)})$
is symmetric in each of variables
$(z_1^{(1)},\cdots,z_l^{(1)})$,
$\cdots, (z_1^{(N)},\cdots,z_l^{(N)})$.
\end{dfn}
The product
$\circ$ is given by the same as (\ref{def:FO3}).
Let us set theta function with parameters
$\nu_1,\cdots,\nu_N$ and
$\alpha$.
\begin{eqnarray}
&&\vartheta_{m,\alpha}^*(u_1^{(1)},\cdots,u_m^{(1)}|\cdots|
u_1^{(N)},\cdots,u_m^{(N)})=
\prod_{t=1}^N
\left[
\sum_{j=1}^m(u_j^{(t+1)}-u_j^{(t)})-\nu_t+\alpha
\right]_{r-k}.
\end{eqnarray}

\begin{prop}~~$\vartheta_{m,\alpha}$ and 
$\vartheta_{n,\beta}$ commute
each other with respect to the product (\ref{def:FO6}).
\begin{eqnarray}
\vartheta^*_{m,\alpha} * 
\vartheta^*_{n,\beta}=
\vartheta^*_{n,\beta} * 
\vartheta^*_{m,\alpha}.
\end{eqnarray}
\end{prop}

\subsection{Free field realization}

In this section we give free field realization of
Level $k$ elliptic algebra $U_{q,p}(\widehat{sl_N})$.
The author would like to emphasize that
the free field realization of Level $k$ is
completely different from
those of Level $1$.
We introduce free bosons 
$a_n^i,(1\leqq i \leqq N-1; n \in {\mathbb Z}_{\neq 0})$,
$b_n^{i,j}, (1\leqq i<j \leqq N; n \in {\mathbb Z}_{\neq 0})$,
$c_n^{i,j}, (1\leqq i<j \leqq N; n \in {\mathbb Z}_{\neq 0})$,
and the zero-mode operators
$a^i,(1\leqq i \leqq N-1)$,
$b^{i,j},(1\leqq i<j \leqq N)$,
$c^{i,j},(1\leqq i<j \leqq N)$.
\begin{eqnarray}
~[a_n^i,a_m^j]&=&\frac{[(k+N)n][A_{i,j}n]}{n}\delta_{n+m,0},~~
[p_a^i,q_a^j]=(k+N) A_{i,j},
\\
~[b_n^{i,j},b_m^{k,l}]&=&-\frac{[n]^2}{n} 
\delta_{i,k}
\delta_{j,l}
\delta_{n+m,0},~~
[p_b^{i,j},q_b^{k,l}]=-\delta_{i,k}\delta_{k,l},
\\
~[c_n^{i,j},c_m^{k,l}]&=&
\frac{[n]^2}{n} \delta_{i,k}\delta_{j,l} \delta_{n+m,0},~~
[p_c^{i,j},q_c^{k,l}]=\delta_{i,k}\delta_{j,l}.
\end{eqnarray}
Here the matrix $(A_{i,j})_{1\leqq i,j \leqq N-1}$
represents the Cartan matrix of classical $sl_N$.
For parameters 
$a_i \in {\mathbb R},(1\leqq i \leqq N-1)$, 
$b_{i,j} \in {\mathbb R},(1\leqq i<j \leqq N)$
$c_{i,j},\in {\mathbb R}, (1\leqq i<j \leqq N)$,
we set
the vacuum vector $|a, b, c\rangle$ of the Fock space 
${\cal F}_{a,b,c}$ as following.
\begin{eqnarray}
&&a_n^i|a,b,c\rangle=b_n^{j,k}|a,b,c\rangle
=c_n^{j,k}|a,b,c\rangle=0,~~(n>0; 1\leqq i \leqq N-1; 
1\leqq j<k \leqq N),\nonumber
\end{eqnarray}
\begin{eqnarray}
p_a^i|a,b,c\rangle=a_i|a,b,c\rangle,~
p_b^{j,k}|a,b,c\rangle=b_{j,k}|a,b,c\rangle,~
p_c^{j,k}|a,b,c\rangle=c_{j,k}|a,b,c\rangle,\nonumber
\\
~~(1\leqq i \leqq N-1; 1\leqq j<k \leqq N).\nonumber
\end{eqnarray}
The Fock space
${\cal F}_{a,b,c}$ 
is generated by bosons $a_{-n}^i,b_{-n}^{j,k},c_{-n}^{j,k}$ 
for $n \in {\mathbb N}_{\neq 0}$.
The dual Fock space
${\cal F}_{a,b,c}^*$ is defined as the same manner. 
In this paper we construct the elliptic analogue of Drinfeld current
for $U_{q,p}(\widehat{sl_N})$ by these bosons $a_n^i, b_n^{j,k}, c_n^{j,k}$
acting on the Fock space.

Let us set the bosonic operators $a_\pm^i(z), a^i(z), (1\leqq i \leqq N-1)$, 
$b_\pm^{i,j}(z), b^{i,j}(z), c^{i.j}(z), (1\leqq i<j \leqq N)$ by
\begin{eqnarray}
a_\pm^i(z)&=&\pm(q-q^{-1})\sum_{n>0}a_{\pm n}^i z^{\mp n} \pm p_a^i {\rm log}q,
\\
b_\pm^{i,j}(z)&=&
\pm(q-q^{-1})\sum_{n>0}b_{\pm n}^{i,j} 
z^{\mp n} \pm p_b^{i,j} {\rm log}q,\\
a^i(z)&=&-\sum_{n \neq 0}\frac{a_n^i}{[(k+N)n]}q^{-\frac{k+N}{2}|n|}z^{-n}+
\frac{1}{k+N}(q_a^i+p_a^i {\rm log}z),\\
b^{i,j}(z)&=&-\sum_{n \neq 0}\frac{b_n^{i,j}}{[n]}z^{-n}+
q_b^{i,j}+p_b^{i,j}{\rm log}z,
\\
c^{i,j}(z)&=&-\sum_{n \neq 0}\frac{c_n^{i,j}}{[n]}z^{-n}+
q_c^{i,j}+p_c^{i,j}{\rm log}z,
\end{eqnarray}
Let us set the auxiliary operators $\gamma^{i,j}(z),\beta_1^{i,j}(z),
\beta_2^{i,j}(z), \beta_3^{i,j}(z), \beta_4^{i,j}(z)$,
$(1\leqq i<j \leqq N)$ by
\begin{eqnarray}
\gamma^{i,j}(z)&=&-\sum_{n \neq 0}\frac{(b+c)_n^{i,j}}{[n]}z^{-n}
+(q_b^{i,j}+q_c^{i,j})+(p_b^{i,j}+p_c^{i,j}){\rm log}(-z),\\
\beta_1^{i,j}(z)&=&b_+^{i,j}(z)-(b^{i,j}+c^{i,j})(qz),
~\beta_2^{i,j}(z)=b_-^{i,j}(z)-(b^{i,j}+c^{i,j})(q^{-1}z),
\\
\beta_3^{i,j}(z)&=&b_+^{i,j}(z)+(b^{i,j}+c^{i,j})(q^{-1}z),~
\beta_4^{i,j}(z)=b_-^{i,j}(z)+(b^{i,j}+c^{i,j})(qz).
\end{eqnarray}
We give a free field realization of Drinfeld current
for $U_q(\widehat{sl_N})$.

\begin{df}~~Let us set the bosonic operators
$E^{\pm,i}(z), (1\leqq i \leqq N-1)$ by
\begin{eqnarray}
E^{+,i}(z)=\frac{-1}{(q-q^{-1})z}\sum_{j=1}^i E^{+,i}_{j}(z),\\
E^{-,i}(z)=\frac{-1}{(q-q^{-1})z}\sum_{j=1}^{N-1} E^{-,i}_{j}(z),
\end{eqnarray}
where we have set
\begin{eqnarray}
E^{+,i}_j(z)=:e^{\gamma^{j,i}(q^{j-1}z)}(
e^{\beta_1^{j,i+1}(q^{j-1}z)}-e^{\beta_2^{j,i+1}(q^{j-1}z)})
e^{\sum_{l=1}^{j-1}(b_+^{l,i+1}(q^{l-1}z)-b_+^{l,i}(q^lz))}:,
\end{eqnarray}
\begin{eqnarray}
E^{-,i}_j(z)&=&:e^{\gamma^{j,i+1}(q^{-(k+j)}z)}
(e^{-\beta_4^{j,i}(q^{-(k+j)}z)}-e^{-\beta_3^{j,i}(q^{-(k+j)}z)})\nonumber\\
&\times& e^{\sum_{l=j+1}^i 
(b_-^{l,i+1}(q^{-(k+l-1)}z)-b_-^{l,i}(q^{-(k+l)}z))+a_-^i(q^{-\frac{k+N}{2}}z)
+\sum_{l=i+1}^N(b_-^{i,l}(q^{-(k+l)}z)-b_-^{i+1,l}(q^{-(k+l-1)}z))}:,
\nonumber\\
&&~~~~~~~~~~~~~~~~~~~{\rm for}~~~1\leqq j \leqq i-1,\\
E^{-,i}_i(z)&=&:e^{\gamma^{i,i+1}(q^{-(k+i)}z)+a_-^i(q^{-\frac{k+N}{2}}z)+
\sum_{l=i+1}^N (b_-^{i,l}(q^{-(k+l)}z)-b_-^{i+1,l}(q^{-(k+l-1)}z))}:\nonumber\\
&-&:e^{\gamma^{i,i+1}(q^{k+i}z)+a_+^i(q^{\frac{k+N}{2}}z)+\sum_{l=i+1}^N
(b_+^{i,l}(q^{k+l}z)-b_+^{i+1,l}(q^{k+l-1}z))}:,\\
E^{-,i}_j(z)&=&:e^{\gamma^{i,j+1}(q^{k+j}z)}
(e^{\beta_2^{i+1,j+1}(q^{k+j}z)}-e^{\beta_1^{i+1,j+1}(q^{k+j}z)})
e^{a_+^i(q^{\frac{k+N}{2}}z)+\sum_{l=j+1}^N (
b_+^{i,l}(q^{k+l}z)-b_+^{i+1,l}(q^{k+l-1}z))}:,\nonumber\\
&&~~~~~~~~~~~~~~~~~~~{\rm for}~~~i+1 \leqq j \leqq N-1.
\end{eqnarray}
Let us set the bosonic operators $\psi_i^\pm(z), (1\leqq i \leqq N-1)$ by
\begin{eqnarray}
\psi_\pm^i(q^{\pm \frac{k}{2}}z)=:e^{\sum_{j=1}^i (b_\pm^{j,i+1}
(q^{\pm (k+j-1)}z)-b_\pm^{j,i}(q^{\pm (k+j)}z))+a_\pm^i(q^{\pm 
\frac{k+N}{2}}z)+\sum_{j=i+1}^N 
(b_\pm^{i,j}(q^{\pm(k+j)}z)-b_\pm^{i+1,j}(q^{\pm (k+j-1)}z))}:.
\end{eqnarray}
Let us set
\begin{eqnarray}
h_i=\sum_{j=1}^i (p_b^{j,i+1}-p_b^{j,i})+p_a^i+\sum_{j=i+1}^N
(p_b^{i,j}-p_b^{i+1,j}).
\end{eqnarray}
\end{df}
Let us introduce the auxiliary operators ${\cal B}_\pm^{* i,j}(z), 
{\cal B}_\pm^{i,j}(z)$,
$(1\leqq i<j \leqq N)$ by
\begin{eqnarray}
{\cal B}_\pm^{* i,j}(z)&=&\exp\left(\pm \sum_{n>0}\frac{1}{[r^*n]}b^{i,j}_{-n}
(q^{r^*-1}z)^{n}\right),
\\
{\cal B}_\pm^{i,j}(z)&=&\exp\left(\pm \sum_{n>0}\frac{1}{[rn]}b^{i,j}_n
(q^{-r^*+1}z)^{-n}\right).
\end{eqnarray}
Let us introduce the auxiliary operators ${\cal A}^{* i}(z), 
{\cal A}^i(z)$, 
$(1\leqq i \leqq N-1)$ by
\begin{eqnarray}
{\cal A}^{* i}(z)&=&
\exp\left(\sum_{n>0}\frac{1}{[r^*n]}a_{-n}^i (q^{r^*}z)^n\right),
\\
{\cal A}^{i}(z)&=&\exp\left(-\sum_{n>0}\frac{1}{[rn]}a_n^i 
(q^{-r^*}z)^{-n}\right).
\end{eqnarray}

\begin{df}~~We define the dressing operators 
${U}^{* i}(z), U^i(z),(1\leqq i \leqq N-1)$.
\begin{eqnarray}
U^{* i}(z)&=&\left(\prod_{j=1}^{i-1}
{\cal B}_+^{* j,i+1}(q^{2-j}z){\cal B}_-^{* j,i}(q^{1-j}z)\right)
\label{def:dress1}\\
&\times&{\cal B}_+^{* i,i+1}(q^{2-i}z){\cal B}_+^{* i,i+1}(q^{-i}z)
\left(\prod_{j=i+2}^N
{\cal B}_+^{* i,j}(q^{-j+1}z){\cal B}_-^{* i+1,j}(q^{-j+2}z)
\right){\cal A}^{* i}(q^{\frac{k-N}{2}}z),\nonumber
\\
{U}^i(z)&=&\left(\prod_{j=1}^{i-1}
{\cal B}_-^{j,i+1}(q^{-2+j}z){\cal B}_+^{j,i}(q^{-1+j}z)\right)
\label{def:dress2}
\\
&\times&{\cal B}_-^{i,i+1}(q^{-2+i}z){\cal B}_-^{i,i+1}(q^{i}z)
\left(\prod_{j=i+2}^N
{\cal B}_-^{i,j}(q^{j-1}z){\cal B}_+^{i+1,j}(q^{j-2}z)
\right){\cal A}^i(q^{\frac{-k+N}{2}}z).\nonumber
\end{eqnarray}
\end{df}

\begin{df}~~
We define the elliptic deformation of Drinfeld current 
$E_i(z), F_i(z), H_i^\pm(z), (1\leqq i \leqq N-1)$, by
\begin{eqnarray}
&&E_i(z)=U^{* i}(z)E^{+,i}(z)e^{2Q_i}z^{-\frac{P_i-1}{r-k}},\\
&&F_i(z)=E^{-,i}(z)U^i(z)z^{\frac{h_i+P_i-1}{r}},\\
&&H_i^+(z)=U^{* i}(q^{\frac{k}{2}}z)\psi_i^+(z)U^i(q^{-\frac{k}{2}}z)
e^{2Q_i}
q^{-h_i}
(q^{(r-\frac{k}{2})}z)^{\frac{h_i+P_i-1}{r}-\frac{P_i-1}{r^*}},\\
&&H_i^-(z)=U^{* i}(q^{-\frac{k}{2}}z)\psi_i^
-(z)U^i(q^{\frac{k}{2}}z)
e^{2Q_i}
q^{h_i}
(q^{-(r-\frac{k}{2})}z)^{\frac{h_i+P_i-1}{r}-\frac{P_i-1}{r^*}}.
\end{eqnarray}
\end{df}

\begin{thm}~~The bosonic operators $E_i(z), F_i(z), H_i^\pm(z)$, 
$(1\leqq i,j \leqq N-1)$
satisfy the following commutation relations.
\begin{eqnarray}
~&&[u_1-u_2-\frac{A_{i,j}}{2}]_{r-k}
E_i(z_1)E_j(z_2)=
[u_1-u_2+\frac{A_{i,j}}{2}]_{r-k}E_j(z_2)E_i(z_1),
\label{eqn:e1}
\\
~&&[u_1-u_2+\frac{A_{i,j}}{2}]_{r}
F_i(z_1)F_j(z_2)=
[u_1-u_2-\frac{A_{i,j}}{2}]_{r}
F_j(z_2)F_i(z_1),
\label{eqn:e2}
\end{eqnarray}
\begin{eqnarray}
~[E_i(z_1),F_j(z_2)]=\frac{\delta_{i,j}}{(q-q^{-1})z_1z_2}\left(
\delta\left(q^{-k}\frac{z_1}{z_2}\right)H_i^+(q^{-\frac{k}{2}}z_1)-
\delta\left(q^{k}\frac{z_1}{z_2}\right)H_i^-(q^{-\frac{k}{2}}z_2)\right).
\label{eqn:e3}
\end{eqnarray}
\end{thm}
We have constructed
the free field realization
of the elliptic algebra $U_{q,p}(\widehat{sl_N})$
for Level $k\neq 0,-N$.
In order to construct free field realiztion of a pair
of Feigin-Odesskii algebra
(\ref{def:FO3}) and (\ref{def:FO6}),
we have to solve the following problem.

~\\
{\bf Problem}~~(1)~Construct the free field realization
of the currents
$E_N(z)$ and $F_N(z)$, which satisfy
the relations (\ref{eqn:e1}), (\ref{eqn:e2}) 
and (\ref{eqn:e3}) which are valid for all
$1 \leqq i,j \leqq N$.~
(2)~Construct one parameter $s$ deformation
of the free field realiztion of
$E_j(z),F_j(z)$, $(1\leqq j \leqq N)$.
\\
\\
After finishing the above problem, it is not 
difficult to construct the free field realization
${\cal G}$ and ${\cal G}^*$ 
of a pair of Feigin-Odesskii algebra
(\ref{def:FO3}) and (\ref{def:FO6}).

\section*{Acknowledgement}~~~
The author would like to thank
the organizing committee of
RIMS Workshop
"Mathematical method of integrable systems 
and its application", 
held in Hakodate, Japan, 2009.
This work is supported by the Grant-in Aid
for Scientific Research {\bf C}(21540228) 
from Japan Society for
Promotion of Science.

\end{document}